\documentclass[conference]{IEEEtran}
\pdfoutput=1
\IEEEoverridecommandlockouts
\usepackage{cite}
\usepackage{amsmath,amssymb,amsfonts}
\usepackage[ruled,vlined]{algorithm2e}
\usepackage{graphicx}
\usepackage{textcomp}
\usepackage{xcolor}
\usepackage{algpseudocode}
\usepackage{balance}

\def\BibTeX{{\rm B\kern-.05em{\sc i\kern-.025em b}\kern-.08em
    T\kern-.1667em\lower.7ex\hbox{E}\kern-.125emX}}
\begin{document}

\title{Bayesian Optimisation for a Biologically Inspired Population Neural Network}

\author{\IEEEauthorblockN{Mahak Kothari}
\IEEEauthorblockA{\textit{Computer Science \& Information Systems} \\
\textit{BITS Pilani, Goa Campus}\\
Goa, India \\
f20180232@goa.bits-pilani.ac.in}
\and
\IEEEauthorblockN{Swapna Sasi}
\IEEEauthorblockA{\textit{Computer Science \& Information Systems} \\
\textit{BITS Pilani, Goa Campus}\\
Goa, India \\
p20190054@goa.bits-pilani.ac.in}
\and
\IEEEauthorblockN{Jun Chen}
\IEEEauthorblockA{\textit{School of Engineering and Materials Science} \\
\textit{Queen Mary University of London}\\
London, United Kingdom \\
jun.chen@qmul.ac.uk}
\and
\IEEEauthorblockN{Elham Zareian}
\IEEEauthorblockA{\textit{School of Engineering} \\
\textit{University of Lincoln}\\
Lincoln, United Kingdom \\
ezareian@lincoln.ac.uk}
\and
\IEEEauthorblockN{Basabdatta Sen Bhattacharya}
\IEEEauthorblockA{\textit{Computer Science \& Information Systems} \\
\textit{BITS Pilani, Goa Campus}\\
Goa, India \\
basabdattab@goa.bits-pilani.ac.in}
}

\maketitle

\begin{abstract}
We have used Bayesian Optimisation (BO) to find hyper-parameters in an existing biologically plausible population neural network. The 8-dimensional optimal hyper-parameter combination should be such that the network dynamics simulate the resting state alpha rhythm (8~--~13 Hz rhythms in brain signals). Each combination of these eight hyper-parameters constitutes a `datapoint' in the parameter space. The best combination of these parameters leads to the neural network's output power spectral peak being constraint within the alpha band. Further, constraints were introduced to the BO algorithm based on qualitative observation of the network output time series, so that high amplitude pseudo-periodic oscillations are removed. Upon successful implementation for alpha band, we further optimised the network to oscillate within the theta (4~--~8 Hz) and beta (13~--~30 Hz) bands. The changing rhythms in the model can now be studied using the identified optimal hyper-parameters for the respective frequency bands. We have previously tuned parameters in the existing neural network by the trial-and-error approach; however, due to time and computational constraints, we could not vary more than three parameters at once. The approach detailed here, allows an automatic hyper-parameter search, producing reliable parameter sets for the network.

\end{abstract}


\section{Introduction}
\label{sec:1}
Population neural networks, also referred to as lumped parameter~\cite{LopesdaSilva1974} or neural mass~\cite{freeman1975mass} models, are used commonly to simulate brain alpha rhythms~\cite{Suffczynski1999}, oscillations at $8-13$ Hz frequency in the scalp Electroencephalogram (EEG) and Local Field Potentials (LFP) of the thalamocortical tissue~\cite{hughes05}, ~\cite{wilsoncowan73}, ~\cite{niedermeyer97}. Increased biological plausibility in such biologically inspired neural networks also results in an inflated parameter space. Tuning such multidimensional parameter space by the trial-and-error approach to obtain the desired frequency domain response is tedious, time-consuming, and is prone to human error. Previously, we have used a Genetic Optimisation algorithm~\cite{elham2016} to tune parameters in a modified version~\cite{Bhattacharya2011} of the original alpha rhythm lumped parameter model~\cite{Suffczynski2004}, that was simulated on Mathwork Simulink~\cite{MATLAB2020}. While our results were encouraging, the run-time was very high even for a small set of parameters on which to optimise the model. Thus, we looked for alternative mechanisms to automate the parameterising of the current version of the alpha rhythm model~\cite{bhattacharya2017} that implements cellular mechanisms of the synaptic pathways~\cite{destexhe98}. In a preliminary testing phase, we have compared the grid search, random search and BO to optimally parameterise an example case; the BO performed better than the other two in terms of finding the minima of an example function. Based on this study, we have used the BO algorithm to find the parameter set of our population neural network such that the model power spectra has a peak within the alpha rhythm. In addition, we have also found the model hyper-parameters that correspond to the model's  response at the theta ($4 - 8$ Hz) and beta bands ($13 - 30$ Hz). These bands are often used to study various brain states in EEG-based studies~\cite{ursino10, basar08}. While a transition from alpha to theta may indicate a `slowing' of the EEG that is often associated with neurological disorders, the beta band is believed to indicate a working state of the brain (as opposed to relaxing as in alpha). A continuous transition of these parameters in our model network ( as identified by the BO algorithm), show a distinct change in time series response; corresponding frequency domain response indicates the transition from theta to alpha to beta band in the output.

Overall, our study has shown that BO is a good candidate to automate hyper-parameter search in the multidimensional parameter space of our existing biologically inspired neural network to simulate brain rhythms. As future work, we will test the plausibility of this approach in a network with larger parameter space. This will be benchmarked with the Genetic Optimisation algorithm that we have been using in our prior works towards the same objective.

A brief overview of the BO algorithm and the population neural network model is provided in Section~\ref{sec:2}. The simulation methods and corresponding results are presented in Section ~\ref{sec:3}. We conclude the paper in section ~\ref{sec:4}. 
\section{Methods}
\label{sec:2}
\subsection{Bayesian Optimisation}
\label{sec:21}
BO adopts a strategy to find the global minimum of a complex non-convex surface using an `exploration-exploitation' technique. In an interesting study on how humans optimise their decision making process in the highly complex environment that they live in, machine optimisations using Gaussian Process (GP) came close to resembling human-like performance~\cite{borji2013}. Thus, the study observed BO, which uses Gaussian Processes as the prior, predicted human search methods more closely than several other existing standard optimisation algorithms in Machine Learning.

We do not try to define the BO algorithm here; interested readers may refer to ~\cite{jones1998} for more details. The three main constituents of BO are: an 'objective function' ($F$) that is to be optimised (finding the global extremum); a set of hyper-parameters $\xi \in \mathcal{R}^D$ ($D$-dimensional space of real numbers $\mathcal{R}$) on which $F$ will be optimised; an acquisition function that is used to decide on the next $\xi$ on which $F$ will be evaluated. The acquisition function decides the trade-off between exploration and exploitation. A high exploitation runs the risk of getting stuck to a local minimum, while a high exploration may lay too much emphasis on global search, and may be equally undesirable~\cite{jones1998}.
Expected Improvement($EI$), an algorithm proposed in~\cite{jones1998} (although note that the author credited~\cite{mockus1978} for the initial idea), is known to provide a good balance of exploration and exploitation in search for the next $\xi$ to obtain the posterior distribution (a prediction of the true objective function). 

Here, we have used our biologically inspired population neural network oscillating within the dominant frequency range of alpha rhythm, and implemented on Matlab, as the objective function (F). In the following section, we introduce the population neural network, and then discuss how BO is applied to find a set of its optimal hyper-parameters.

\subsection{The Population Neural Network}
\label{sec:22}
\begin{figure}[ht]
\begin{center}
\includegraphics[width=0.49\textwidth]{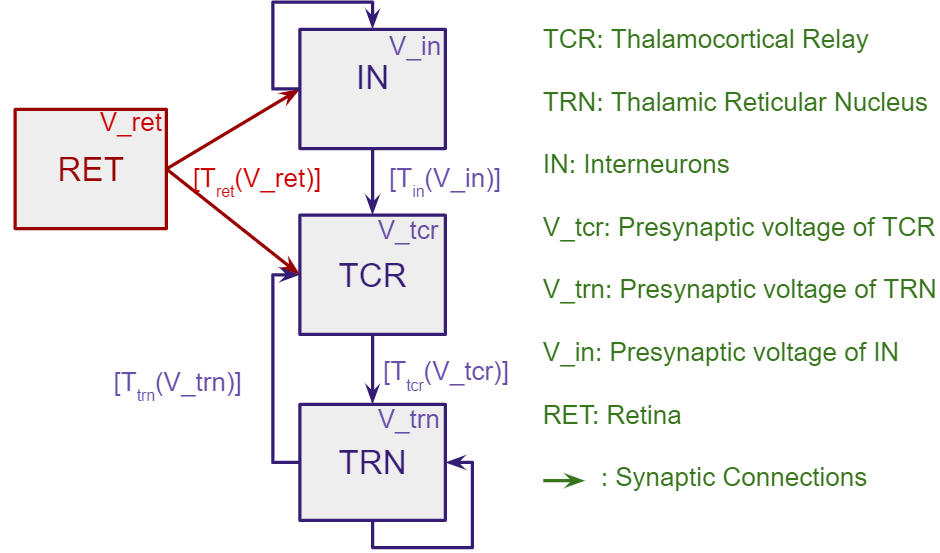}
\caption{Synaptic Layout of Lateral Geniculate Nucleus (LGN), which is the part of the thalamus that process sensory information from the eyes. The LGN has an excitatory Thalamocortical Relay (TCR) population that are the main carriers of sensory information from the eyes to the visual cortex. There are the inhibitory local interneuron population (IN); both IN and TCR receive sensory input from the Retinal neuron population (RET). In addition, there are the inhibitory Thalamic Reticular Nucleus (TRN) population that surround the LGN and communicates with the TCR bidirectionally. We use BO on this model to find an optimal hyper-parameter set such that the LGN network output (the TCR response) oscillates with a dominant frequency within the alpha (8~--~13 Hz band). These hyper-parameters in our work are the `bias' ($ V_{pre} $) and `slope'($ \sigma $) parameters of the sigmoid functions that define the neurotransmitter concentration in the synaptic cleft corresponding to each neural population in the network ($[T (V_{pre})]$); please refer to Eq.~\ref{eq:1}. As there are 4 cell populations, there will be four such sigmoids, each of which will have two hyper-parameters. Thus, we have a total of 8 hyper-parameters on which the BO algorithm is used to find the values for the LGN model.}
\label{fig:LGN}
\end{center}
\end{figure}

The Thalamus forms the relay centre in the brain from where all (except olfactory) sensory information is conveyed to the `deepest' (with reference to the sensory organs) part of the brain, viz. the neocortex. We have been working with a population neural network of the thalamic `nucleus' that forms the visual pathway, known as the Lateral Geniculate Nucleus (LGN)~\cite{Bhattacharya2011}. The LGN circuit has three connected neural populations as shown in Fig.~\ref{fig:LGN}. Every connection shown in the network models a synaptic connectivity from a pre-synaptic (`from') population to a post-synaptic (`to') population. Extrinsic input to the network is white noise simulating noisy inputs from the retina. 

Subsequently, more biologically inspired parameters were introduced in the model with a goal to better understand the synaptic dynamics underlying altered brain rhythm~\cite{Bhattacharya2013}. However, such increased attributes led to a non-linear increase in the parameter space; each synaptic pathway in this model has 8 hyper-parameters, in contrast to 3 in the previous version. We have used a BO algorithm based approach on this recent variant of the LGN population neural network, focusing on parameterising the dynamics of the neuro-transmitter concentration in the synaptic cleft.

The neurotransmitter released in the synaptic cleft by a pre-synaptic population is a function of its voltage, and is defined in~(Eq.~\ref{eq:1}):\\

\begin{equation} \label{eq:1}
    [T_{pre}(V_{pre})] = \frac{T_{max}}{1+e^{-\frac{V{pre} - V_{thr}}{\sigma}}},
\end{equation}
where $[T_{pre}]$ represents the concentration of neurotransmitters in the synaptic cleft corresponding to the population $\zeta_{pre}: \zeta \in \{TCR, TRN, IN, RET\}$, $T_{max}$ represents maximum neurotransmitter concentration and 1mM is considered to be a biologically plausible value~\cite{destexhe1994}, $V_{thr}$ is the threshold voltage at which $[T_{pre}]$ becomes half of $T_{max}$ and $\sigma$ affects the steepness of this sigmoid function. Physiological studies do not provide exact values of the threshold voltage $V_{thr}$ nor that of the steepness parameter $ \sigma $ of the sigmoid function that defines the dynamics of the neurotransmitters in the synaptic cleft. Thus, there are a total of 8 hyper-parameters corresponding to each of the four populations in the network (see Fig.~\ref{fig:LGN}).

In our previous works~\cite{bhattacharya2017}, we have set these values by trial and error to $\approx [(\mbox{resting membrane potential}) + (30 \pm \epsilon)]$ mV, where $\epsilon$ is an arbitrarily selected range that is tuned to obtain desirable model dynamics; resting membrane potential is -65mV. In a similar manner, the existing model has $3 \leq \sigma \leq 4$ tuned by trial and error. However, an eight dimensional parameter space is hard to parameterise manually. Thus, all $[T_{pre}]$ in the current LGN have the same values; $V_{thr} = -32mV $ and $\sigma = 3.8 $mV for all connections. This is one of the main drawbacks in our existing model. Readers may note that the synaptic connection weights in the model are in the biologically plausible range (see~\cite{Bhattacharya2011} for details); in addition, over the course of several works, they are tuned to values that produce a dominant frequency oscillation within the alpha band for the network with all parameters set to base values~\cite{bhattacharya2017}. Thus, we do not consider the synaptic weights as hyper-parameters for optimising.
\subsection{Applying BO to the Network}
\label{sec:23}
With this approach based on BO algorithm, the population neural network of LGN, $V_{thr} $ and $\sigma $ are varied over the ranges $-40$ to $-28$ mV and $3$ to $5$ mV respectively for all four neural populations, thus making it possible to optimise the network over eight hyper-parameters at once. The datapoints $\xi \in \mathcal{R}^D$ are each set of eight hyper-parameters that form a vector in $\mathcal{R}^8$, and selected from the afore-mentioned ranges; readers may refer to Fig.~\ref{fig:LGN} legend. Relating to the BO description in Section~\ref{sec:21}, the objective function $F$ is the LGN model output constrained by the alpha rhythmic bound in the frequency domain. 

In this work, we have used the \textit{bayesopt} in-built function in Matlab to implement BO. Other than the objective function $F$ and the set of hyper-parameters specifying $\xi$, the user needs to specify the desired acquisition function to \textit{bayesopt}. A plethora of acquisition functions are available to implement in Matlab. We have chosen the 'Expected-improvement-plus', which is based on the $EI$ algorithm~\cite{jones1998} mentioned in Section~\ref{sec:21}; the 'plus' in the nomenclature is specific to the Matlab utility and implies a further modification on the traditional EI algorithm to deter over-exploitation (please refer to the Matlab documentation). Furthermore, we have to provide the number of trials $N$ over which the algorithm should scan the combinations for $\xi$ to obtain the extremum. This is a hyper-parameter of the \textit{bayesopt} function and was set by trial-and-error in our implementation as $N=200$. At the end of each trial run, the algorithm is rewarded/punished based on how well the returned $\xi$ minimises the error, i.e.\ satisfies the constraint of the peak spectral power ($\mathcal{M}_{peak}$) of the network output time-series satisfying the constraint $8Hz \leqslant \mathcal Freq({M}_{peak}) \leqslant 13Hz$. The procedure is mentioned below.

Since the \textit{bayesopt} function tries to minimize the return value at each trial, we choose the return value as $-\mathcal{M}_{peak}$. Higher negative values of $\mathcal{M}_{peak}$ are used to reward the algorithm, while an arbitrary positive values are chosen as the penalty. In general, the reward is $-\mathcal{M}_{peak}$, if the highest peak of the power spectral density (psd), i.e.\ the first peak in a multi-modal psd is within the given band ( along with additional constraints ). The penalty system used here is hierarchical, with the first unsatisfied rule deciding the penalty. So, an arbitrary positive value like 2 or 1 is used as penalty.  This ensures that a greater reward is awarded if $\xi$ resulted in peak power within the band. \\

For alpha band, the algorithm is rewarded with $-\mathcal{M}_{peak}$ if it satisfies both the below rules:\\
Rule1: the peak spectral power is bounded thus: $8Hz \leqslant \mathcal Freq({M}_{peak}) \leqslant 13Hz$. \\
Rule2: the peak-to-peak voltage of TCR ($V\_tcr\_p2p$) is bounded thus: $0.26$ mV $\leqslant V\_tcr\_p2p \leqslant 0.30$ mV. 

The penalty factors are listed below in decreasing order of severity for alpha band:\\
1) If Rule1 is unsatisfied, then an arbitrary high positive value is returned. In our case, we set this value to 2.\\
2) If Rule2 is unsatisfied, then an arbitrary lower positive value is returned (as compared to Rule1). We set this to 1.

After the total trial run of $N=200$, the \textit{bayesopt} function selects the $\xi$ that gives the highest value $\mathcal{M}_{peak}$ within the above-mentioned bounds for alpha rhythm. 

Following a successful implementation of the BO algorithm on our network, whereby we obtained a robust set of hyper-parameters that was fairly invariant to noisy inputs (see Section~\ref{sec:3} for details of implementation) and produced consistent alpha rhythm, we explored the extension of the model for theta and beta bands. This required customisation of the penalty/reward factors in order to achieve the desired band. The algorithm is rewarded with $-\mathcal{M}_{peak} \pm V\_tcr\_p2p$ (plus for beta, minus for theta) provided it satisfies both the below rules:\\
Rule1: $4Hz \leqslant \mathcal Freq({M}_{peak}) < 8 Hz$ for theta; $13 Hz < \mathcal Freq({M}_{peak}) \leqslant 30 Hz$ for beta. \\
Rule2: $V\_tcr\_p2p > 0.30$ mV for theta; $V\_tcr\_p2p < 0.26$ mV for beta. 

The penalty for theta and beta follows same rule as that for alpha mentioned above. Further details of the algorithm are mentioned in the following section.

\section{Results}
\label{sec:3}
Our initial requirement to obtain the model's peak power spectrum within the alpha band led us to introduce the first constraint and the related reward/penalty factors as described in the previous section. However, due to the stochastic nature of the network input, we did not get consistent alpha rhythm output with the returned best value of $\xi$ by the BO algorithm. Fig. 2 shows an example of such inconsistency corresponding to $\xi$=Step 1 in Table~\ref{tab:2}. 
\begin{figure}[ht]
\begin{center}
 \includegraphics[width=0.5\textwidth]{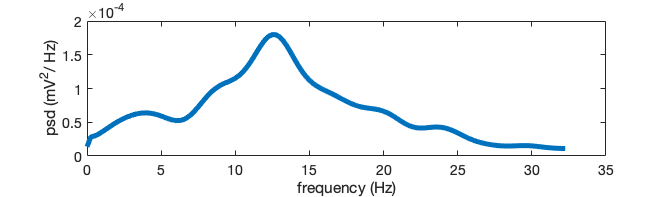}
 \label{fig:alpha_2_1}
 \includegraphics[width=0.5\textwidth]{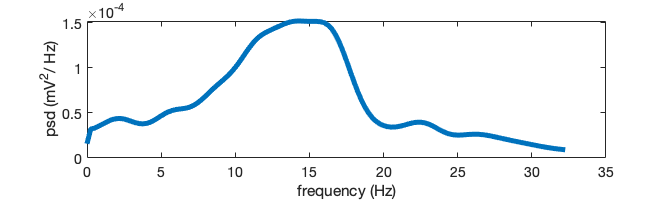}
 \label{fig:alpha_2_2}
\caption{Power Spectral Density corresponding to Step 1 in Table~\ref{tab:2}: (top) Alpha); (bottom) not Alpha.}
\end{center}
\end{figure}

This required us to put further constraints on our system. In the following sections, we discuss the evolution process of the constraints, related algorithms and our findings. 
\subsection{Constraint for Noise Input}
\label{sec:31}
\begin{figure}[ht]
\begin{center}
\includegraphics[width= 0.48\textwidth]{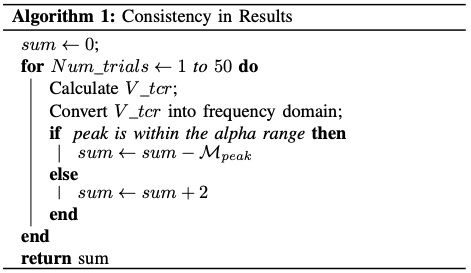}
\caption{Algorithm 1: Constraint to correct for stochastic input to the LGN network which comprise the objective function to the BO algorithm.}
\label{fig:Alg1}
\end{center}
\end{figure}

To  ensure the algorithm returned a robust parameter set that gives the alpha rhythm consistently, we introduced an improvement to our base algorithm. This is shown in Fig.~\ref{fig:Alg1}. The LGN network that forms the core of the objective function to \textit{bayesopt} is simulated for an arbitrary `t=50' independent iterations corresponding to each $\xi$ obtained during one trial run (out of N) of the BO algorithm. In each iteration $t$, the reward (penalty) obtained are subtracted (added) to previous iteration's values. (Reward corresponds to subtraction because we want to get the minimum value). This leads to a cumulative reward/penalty factor at the end of the 50 iterations. Note, the reward/penalty values are reset (\emph{sum}=0 in Fig.~\ref{fig:Alg1}) prior to next trial of the BO algorithm. With this change, the inconsistency in the alpha peak is resolved and an optimal hyper-parameter set giving consistent peak power in alpha band was produced by the algorithm (see Table~\ref{tab:2} step 2).
\subsection{Constraint for Network Output Time-series} 
\label{sec:32}
\begin{table*}[t]
\centering
\caption{Stepwise Parameters for Different Frequency Bands}
\label{tab:2}
\begin{tabular}{|c|c|c|c|c|c|c|c|c|c|}
 \hline
 \textbf{Step} & \textbf{Frequency band} & \textbf{Vp} & \textbf{Kp} & \textbf{Vp\_tcr} & \textbf{Kp\_tcr} & \textbf{Vp\_trn} & \textbf{Kp\_trn} & \textbf{Vp\_in} & \textbf{Kp\_in}  \\
 \hline
  1 & Alpha & -33.131043 & 3.771233 & -38.625510 & 4.474605 & -30.332203 & 4.449851 & -31.869748 & 3.641744 \\
  \hline
  2 & Theta & -29.629145 & 3.092783 & -29.554469 & 4.507881 & -35.637984 & 4.015976 & -34.958183 & 4.017226\\
  \hline
  2 & Alpha & -30.7253 & 3.7915 & -37.4431 & 3.8726 & -31.1964 & 3.9395 & -31.7612 & 3.3257 \\
  \hline
  2 & Beta & -37.857715 & 3.840154 & -38.781024 & 3.537155 & -33.948101 & 3.273600 & -28.795152 & 3.298834\\
  \hline
  3 & Theta & -28.8873 & 3.7727 & -35.3013 & 3.0633 & -29.4486 & 3.1992 & -28.0839 & 3.0993\\
  \hline
  3 & Alpha & -30.9197 & 3.9349 &  -32.8924 & 4.2546 & -38.5436 & 3.4323 & -33.4399 & 3.0939\\
  \hline
  3 & Beta & -29.4522 & 4.0306 & -38.8018 & 4.8982 & -31.9666 & 3.1592 & -37.4803 & 3.0655\\
  \hline
\end{tabular}
\end{table*}
Next, we tested the algorithm for finding a consistent hyper-parameter space for network output within theta and beta bands. Though initial results looked promising, an analysis of the time series plot for each frequency band (Fig. 4) showed discrepancies. It is well known that the peak-to-peak voltage in scalp EEG is low for higher frequencies such as beta, and increases progressively with slower frequencies. Thus, we expected a higher peak-to-peak voltage for alpha oscillations compared to beta~\cite{niedermeyer1983}. However, certain $\xi$ in the network results in the reverse case. This was not a phenomenon that we were familiar with in our previous work, which may be attributed to our fixed values for a standard [$T_{pre}$] model for all populations in the network.
\begin{figure}[ht]
\begin{center}
\includegraphics[width= 0.5\textwidth]{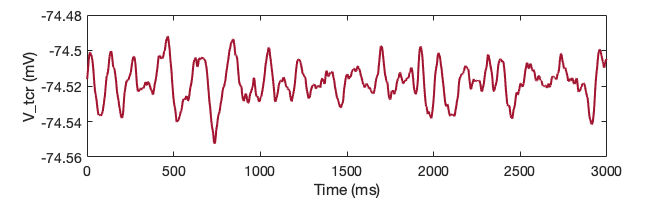}
\label{fig:theta_3} 
\includegraphics[width= 0.5\textwidth]{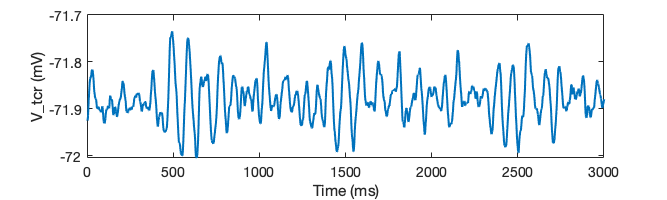}
\label{fig:alpha_3} 
\includegraphics[width= 0.5\textwidth]{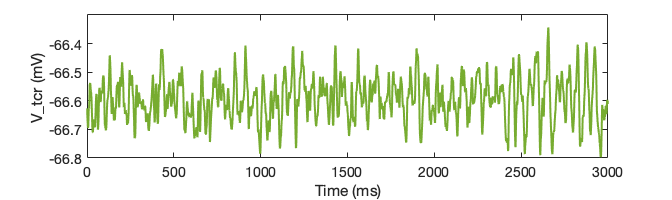}
\label{fig:beta_3} 
\caption{Step 2: Time series corresponding to (top) theta, (middle) alpha, and (bottom) beta oscillations in the model output when simulated using $\xi$ in Step 2 of the Table~\ref{tab:2}.}

\end{center}
\end{figure}
\begin{figure}[ht]
\begin{center}
\includegraphics[width= 0.48\textwidth]{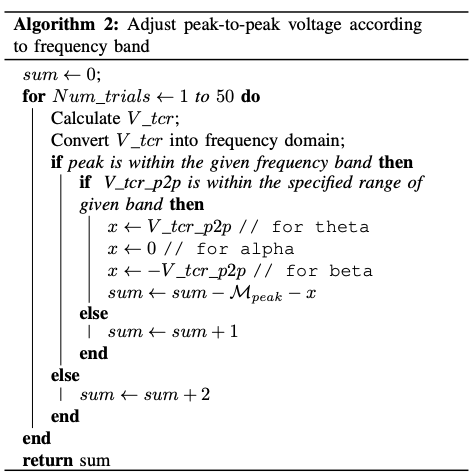}
\caption{Algorithm 2: Constraint to correct for time-series behaviour inconsistencies corresponding to the optimal hyper-parameter set selection by the BO algorithm. }
\label{fig:Alg2}
\end{center}
\end{figure}

As shown in Fig. 4(b), the peak-to-peak voltage of TCR ($V\_tcr\_p2p$) population for alpha band is observed to be around $0.28$ mV. We base the bounds of $V\_tcr\_p2p$ corresponding to theta and beta bands based on this. The Algorithm implementing the constraint is shown in Fig.~\ref{fig:Alg2}. A new component \emph{x} is introduced for deciding the reward and penalty and is set as follows:\\
1) for the alpha band, if $0.26 mV \le V\_tcr\_p2p \le 0.30mV$ , then model is rewarded with minus $\mathcal{M}_{peak}$ else, penalized with arbitrary positive value like +1. In this band,  the new component \emph{x} is unused and hence set to 0 indicating no change to reward value.\\
2) for theta band, if V\_tcr\_p2p $>$ 0.30mV , then \emph{x} is set to V\_tcr\_p2p . Thus, to achieve more reward, the algorithm has to strive to increase the maximum amplitude while decreasing the minimum amplitude of V\_tcr in each trial of BO.\\
3) for the beta band, if V\_tcr\_p2p $<$ 0.26mV, then \emph{x} is set to minus V\_tcr\_p2p. So, in this case to achieve more reward, the algorithm will strive to decrease the maximum amplitude while increasing the minimum amplitude of V\_tcr in each trial of BO.

Applying the algorithm in Fig.~\ref{fig:Alg2}, we could achieve the optimal set of parameters for theta and beta band oscillations. 

To test the robustness of the best hyper-parameter set given out by the algorithm ( refer Table~\ref{tab:2} against "Step 3".), we ran 1000 trials of the independent model for each band. We had 100\% requirement satisfaction for theta and beta band while approximately 65\% for alpha band. The time series and psd plots for each band is shown in Fig. 6 demonstrate our observations.
\begin{figure}[ht]
\centering
\label{fig:theta_uf_4} \includegraphics[width= 0.5\textwidth]{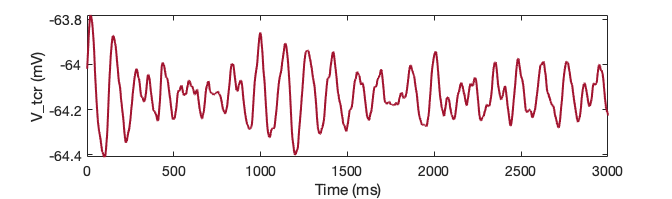}
 \label{fig:theta_4} \includegraphics[width=0.5\textwidth]{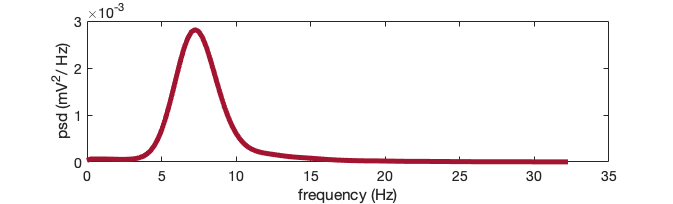}
 \label{fig:alpha_uf_4} \includegraphics[width= 0.5\textwidth]{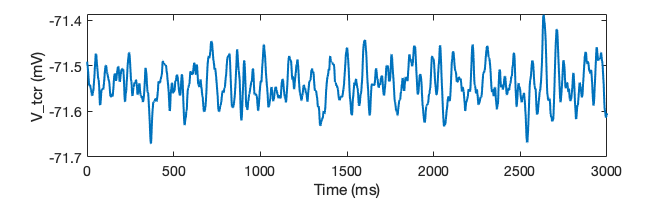}
\label{fig:alpha_4} \includegraphics[width=0.5\textwidth]{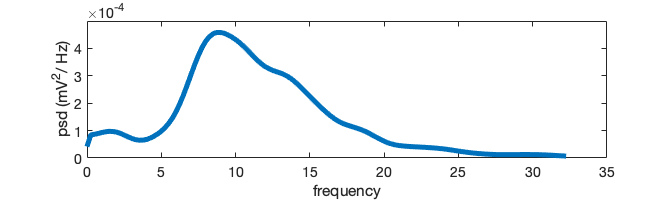}
\label{fig:beta_uf_4} \includegraphics[width= 0.5\textwidth]{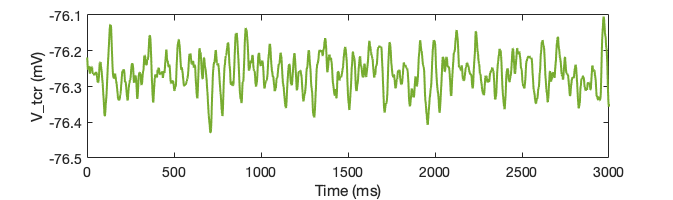}
\label{fig:beta_4} \includegraphics[width=0.5\textwidth]{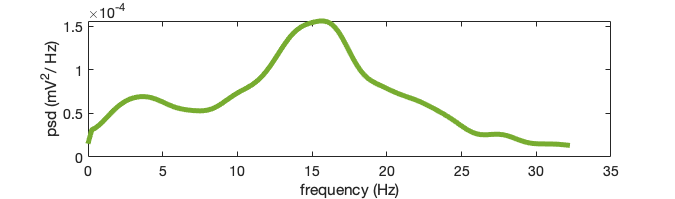}
\caption{Time series and PSD of the network output upon application of Algorithm 2 (Fig.~\ref{fig:Alg2}), and corresponding to $\xi$ shown in Step 3 of Table~\ref{tab:2} for Panels 1 (Top) and 2: Theta, Panels 3 and 4: Alpha, Panels 5 and 6 (bottom): Beta.}
\end{figure}
\label{sec:33}
\begin{figure*}[t]
\begin{center}
\includegraphics[scale=0.7]{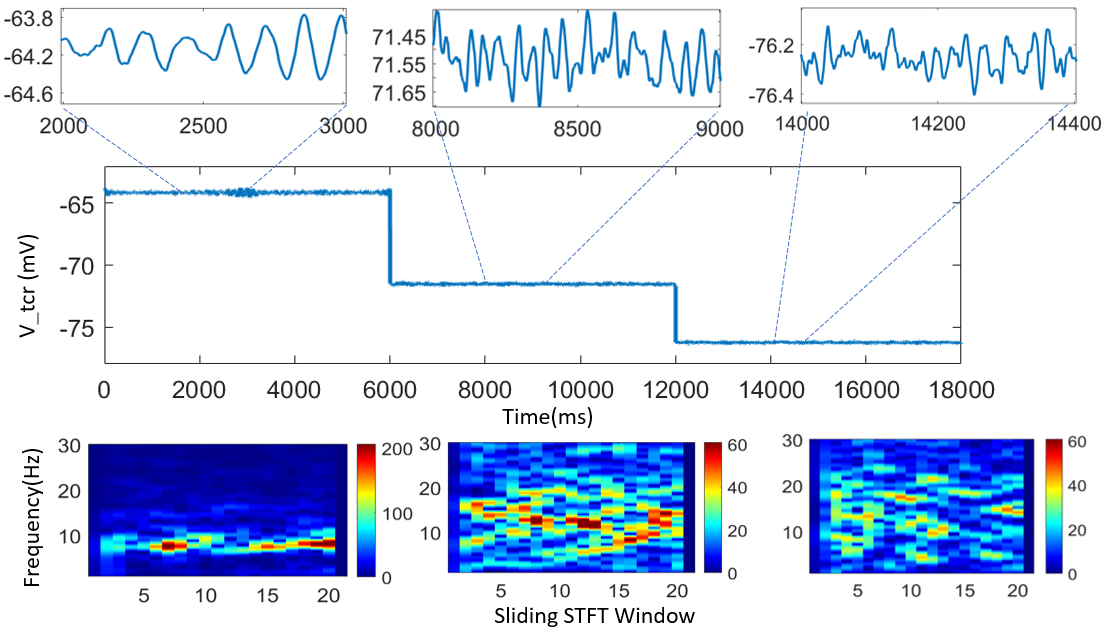}
\caption{Time series and Frequency domain response of the network output corresponding to dynamic change in the hyper-parameter sets that were identified by the BO algorithm as the best for response within the theta, alpha and beta bands. The mean membrane voltage progressively decreases from theta to beta as expected. Zoomed in plots show progressively increased frequency content, and lower peak-to-peak voltage. The Short-Time-Fourier-Transform heat maps show distinct theta and alpha band in the network output. The response to beta band is noisy, which is not surprising for a network that do not model the cortical dynamics.}
\label{fig:stft}
\end{center}
\end{figure*}
\subsection{Continuous time-Series}
Having obtained a reliable hyper-parameter sets for the network output to oscillate within all three frequency bands viz.\ theta, alpha and beta, we set out to simulate a continuous output of the network corresponding to dynamic variation of the hyper-parameters. Thus, when transitioning from the hyper-parameter set corresponding to theta, the network variable states would not be reset. This would further test the robustness of the parameter sets given by the BO algorithm for the network. In biology, it is well known that cellular and synaptic attributes change with changing brain states, which are reflected by different brain rhythms. Thus, our experimental set-up here will simulate real-time change in brain states with the population neural network model of LGN.

To set up our simulation experiment, we first generated a white noise input signal of 21 seconds at a resolution of 1 millisecond. To activate the hyper-parameters dynamically, a timeslot of 7 seconds from the 21 seconds duration is dedicated to theta ($0-7$seconds), alpha ($7-14$ seconds) and beta ($14-21$ seconds). Within each slot, the hyper-parameters are changed to meet the respective band requirements. Initial 1 second in each slot was removed to drop the transients in the time series for the analysis. A zoomed in view of the time series in Fig.~\ref{fig:stft} indicate the peak-to-peak amplitude of V\_tcr is met for the respective band as seen in the individual cases. In addition, a frequency domain analysis is done using Short-Time Fourier Transform (STFT) with sliding window of 1000 milliseconds and 25\% overlap between the windows. Our results show a high power (indicated by orange to red colour in Fig.~\ref{fig:stft}) in the corresponding frequency range of the bands. Of these, alpha and theta bands are relatively higher than other frequency components in the signal. The beta band is noisy, which is expected, as beta rhythms are mainly attributed to the cortical populations, which are absent in our model. Overall, we can comment that the final set of hyper-parameters produced by the BO algorithm is optimal as well as reliable for use in the LGN network.
\section{Conclusion}
\label{sec:4}
We have used BO to tune hyper-parameters in a population neural network, commonly known as the neural mass models, under the constraint that the frequency domain response of the model is within the alpha ($8-13$ Hz) band. The model is an adaptation of the Alpha Rhythm model proposed during the 1970s, and subsequently the subject of many researches. However, the model was computationally slow as far as we observed in our work on the model. In addition, physiological data on which to parameterise these biologically inspired models are hard to get; where there are data, there is a lot of ambiguity in literature. This led us to look to optimisation algorithms that could automate the parameter search in these networks such that their outputs are functionally similar to our objectives. We have previously applied Genetic Algorithm (GA) to optimise a version of the Alpha Rhythm model, where our hyper-parameters were the synaptic connection weights. However our application was computationally expensive. In recent years, we have adapted the original alpha rhythm model to include synaptic mechanisms, that not only introduce greater biological plausibility, but also improve the computational approximately by an order of 10. This in spite an increase in the parameter space of the model due to the synaptic attributes. Thus, it is much harder to tune the parameter space of the model by trial-and-error. Owing to these, thus far, we have been using a single neurotransmitter concentration function, modelled by a sigmoid, to describe the dynamics for all populations and for all synapses. In fact, this aspect was criticised as one of the primary drawbacks of our existing model. We have addressed this aspect in this work.

We have used the inbuilt \textit{Bayesopt} function in Matlab to find the hyper-parameters defining the neurotransmitter concentration for each population in the network. There were a total of eight hyper-parameters with biological plausible ranges that were provided as the parameter space to the algorithm. While the nuances of the BO algorithm are hidden from the users in Matlab, the acquisition function can be chosen by the user. We have chosen the `Expected Improvement plus' function, that is described (Matlab documentation) to be efficient in balancing the exploration and exploitation in the algorithm. Additional constraints were added to the network so that the model output was within the desired bounds of the alpha rhythm. On successful optimisation of the network for alpha rhythm, we also did the same, now for theta ($4-8$ Hz) and beta rhythms ($13-30$ Hz). The successful identification of the hyper-parameters for each of the three different frequency bands by the BO algorithm was demonstrated with progressive change of hyperparameter values in the model that produced corresponding rhythmic dynamics.

In summary, we have introduced a novel approach based on BO algorithm to find hyper-parameters in neural mass models. The main drawback in our work is a lack of benchmark with other optimisation algorithms. In particular, we would like to benchmark the BO with the GA algorithm used in our previous work. In addition, we would like to explore the other acquisition functions available in Matlab, and compare with the current EI-plus version used in this work. The proof-of-concept study presented here provides an impetus to progressing work on large-scale biologically inspired neural networks with hard-to-tune large parameter spaces.
\newpage
\bibliographystyle{ieeetr}
\balance
\bibliography{bibliography.bib}
\end{document}